\documentclass[aps,prb,twocolumn,amsmath,amssymb,superscriptaddress]{revtex4-2}
\usepackage{graphicx}
\usepackage{dcolumn}
\usepackage{bm}
\usepackage{hyperref}
\usepackage{overpic}
\usepackage{float}
\usepackage[utf8]{inputenc}

\hypersetup{
    colorlinks=true,
    linkcolor=blue,
    citecolor=blue,
    urlcolor=blue
}

\begin{document}

\title{Disorder-induced crossover from phase-averaging to mode-mixing regimes in magnetic domain walls of a second-order topological insulator}

\author{Dong Zhou}
\affiliation{School of Physics and Technology, Nanjing Normal University, Nanjing 210023, China}

\author{Zhe Hou}
\email{zhe.hou@nnu.edu.cn}
\affiliation{School of Physics and Technology, Nanjing Normal University, Nanjing 210023, China}

\date{\today}

\begin{abstract}
We investigate electronic transport across a magnetic domain wall (DW) in a three-dimensional (3D) second-order topological insulator subject to Anderson disorder. In the clean limit, the DW hosts two co-propagating one-dimensional (1D) topological edge states that act as the two arms of an effective Aharonov-Bohm (AB) interferometer, inducing a sinusoidal conductance oscillation. Upon the introduction of disorder, the AB oscillations are suppressed, while a half-quantized plateau of $0.5 e^2/h$ for the ensemble-averaged conductance emerges. Notably, within this plateau, the conductance fluctuation exhibits a distinctive two-step plateau structure, with values of $\sim 0.35  e^2/h$ at moderate disorder, followed by a second plateau at $\sim0.29  e^2/h$ under strong disorder. By developing theoretical frameworks that account for the random-phase interference and inter-mode mixing of the two arms, we identify the first fluctuation plateau as a signature of the phase-averaging regime (PAR) and the second as a signature of the mode-mixing regime (MMR). Furthermore, we show that, in the PAR the conductance follows a U-shaped beta distribution, while it evolves into a uniform distribution in the MMR. The Fano factor associated with shot noise is also computed, which exhibits a similar two-step plateau structure at $1/4$ and $1/3$, corresponding to the PAR and MMR, respectively. Our work provides a clear demonstration of the disorder-induced crossover from PAR to MMR, and highlights the crucial role of second-order conductance cumulants in identifying these transport regimes. The results suggest disorder-engineering as a powerful route for controlling electronic transport across DW-based devices.
\end{abstract}

\maketitle

\section{INTRODUCTION}
\label{sec_intro}
Conductance plateaus (CPs) are hallmarks of numerous profound quantum phenomena in condensed matter physics. A fundamental example is the quantized Hall CP in the quantum Hall (QH) effects~\cite{Klitzing1980New}, whose discovery inspired the broader study of topology in physics~\cite{Thouless1982Quantized, Laughlin1981Quantized}. This subsequently gave rise to the field of topological insulators (TIs), including quantum anomalous Hall insulators (QAHIs)~\cite{Haldane1988Model, Yu2010Quantized, Chang2013Experimental}, quantum spin Hall insulators~\cite{Kane2005Quantum, Kane2005Z2, Bernevig2006Quantum, Konig2007Quantum}, and three-dimensional (3D) TIs~\cite{Fu2007Topological, Chen2009Experimental, Zhang2009Topological, Xia2009Observation, Qi2011Topological}. Similarly, fractional quantized Hall CPs, which are signatures of fractional QH effects or fractional QAH effects~\cite{Tsui1982Two, Cai2023Signatures, Park2023Observation, Xu2023Observation, Lu2024Fractional}, reveal deep connections between strong electron-electron interactions and topology. Beyond topologically nontrivial systems, ballistic transport in ultra-clean mesoscopic systems~\cite{Buttiker1985Generalized, Datta1995Mesoscopic}, such as quantum point contacts~\cite{Wees1988Quantized, Wharam1988One}, yields quantized longitudinal CPs, with the quantization value corresponding to the number of reflectionless propagating channels.

In addition to topological and ballistic mechanisms, disorder serves as another critical factor in inducing such CPs. Among these, the half-quantized CP at $0.5 G_0$ has garnered significant attention, with $G_0$ representing the conductance quantum ($2e^2/h$ for spin-degenerate systems or $e^2/h$ for non-degenerate ones). For instance, in graphene $p-n$ junctions within the QH regime~\cite{WilliamsQuantum2007, AbaninQuantized2007, OzyilmazElectronic2007, KlimovEdge2015, LiDisorder2008, LongDisorder2008, LowBallistic2009, SchmidtMixing2013, MachidaEdge2015, MatsuoEdge2015, KumadaShot2015, DaiMode2017}, when the zeroth electron and hole Landau levels are occupied in their respective electrodes, the QH edge states meet at the interface and undergo complete mixing due to disorder. Consequently, the average transmission probability across the junction becomes 1/2, resulting in the half-quantized CP. Similar electron partitioning processes happen at the domain walls (DWs) of disordered TI thin films~\cite{LongScaling2025} where chiral topological QAH edge states get scattered at the interface, as well as in disordered QAHI-superconductor-QAHI heterojunctions~\cite{HuangDisorder2018} where the QAH edge states percolate into the disordered superconducting region. These processes lead to strong inter-channel mixing, and are thus classified under the mode-mixing regime (MMR). The emergence of CPs in this regime is explained by random matrix theory (RMT) applied to quantum chaotic cavities~\cite{AbaninQuantized2007, BarangeriMesoscopic1994, SavinShot2006, BeenakkerRandom1997}.

Alternatively, there exists another disorder-dependent mechanism for inducing the half-quantizd CP, which arises from the Aharnov-Bohm (AB) effect~\cite{Datta1995Mesoscopic,Washburn1986Aharonov,Webb1988Quantum}. When disorder is introduced into the two arms of an AB interferometer, the dynamical phase difference $\Delta \varphi$ between the arms becomes randomized, which smears out the conductance oscillations, inducing a CP. This process can be exemplified by a perfect AB interferometer with a conductance relation: $G =  G_0 [0.5 + 0.5 \cos{(\Phi + \Delta \varphi)} ] $, where $\Phi = \pi \Phi_{\rm B} /\Phi_0$ is the reduced magnetic flux with $\Phi_{\rm B}$ the real magnetic flux, and $\Phi_0 =h/2e$ is the flux quantum. In the presence of disorder, $\Delta \varphi$ is assumed to be uniformly distributed within the interval $[0, 2\pi]$. The half-quantized CP is then obtained by performing the following phase average:
\begin{align}
\langle G \rangle = \int_{0}^{2\pi} G(\Phi + \Delta \varphi ) {\rm d} \Delta \varphi   =0.5 G_0 .
\end{align}
We refer to this mechanism as the {\it phase-averaging regime} (PAR). To date, the PAR and MMR in disordered systems are discussed separately, whereas a single platform for exploring both regimes within a unified framework remains elusive.  

% === Figure 1 ===
\begin{figure}[t]
    \centering
    \includegraphics[width=\columnwidth]{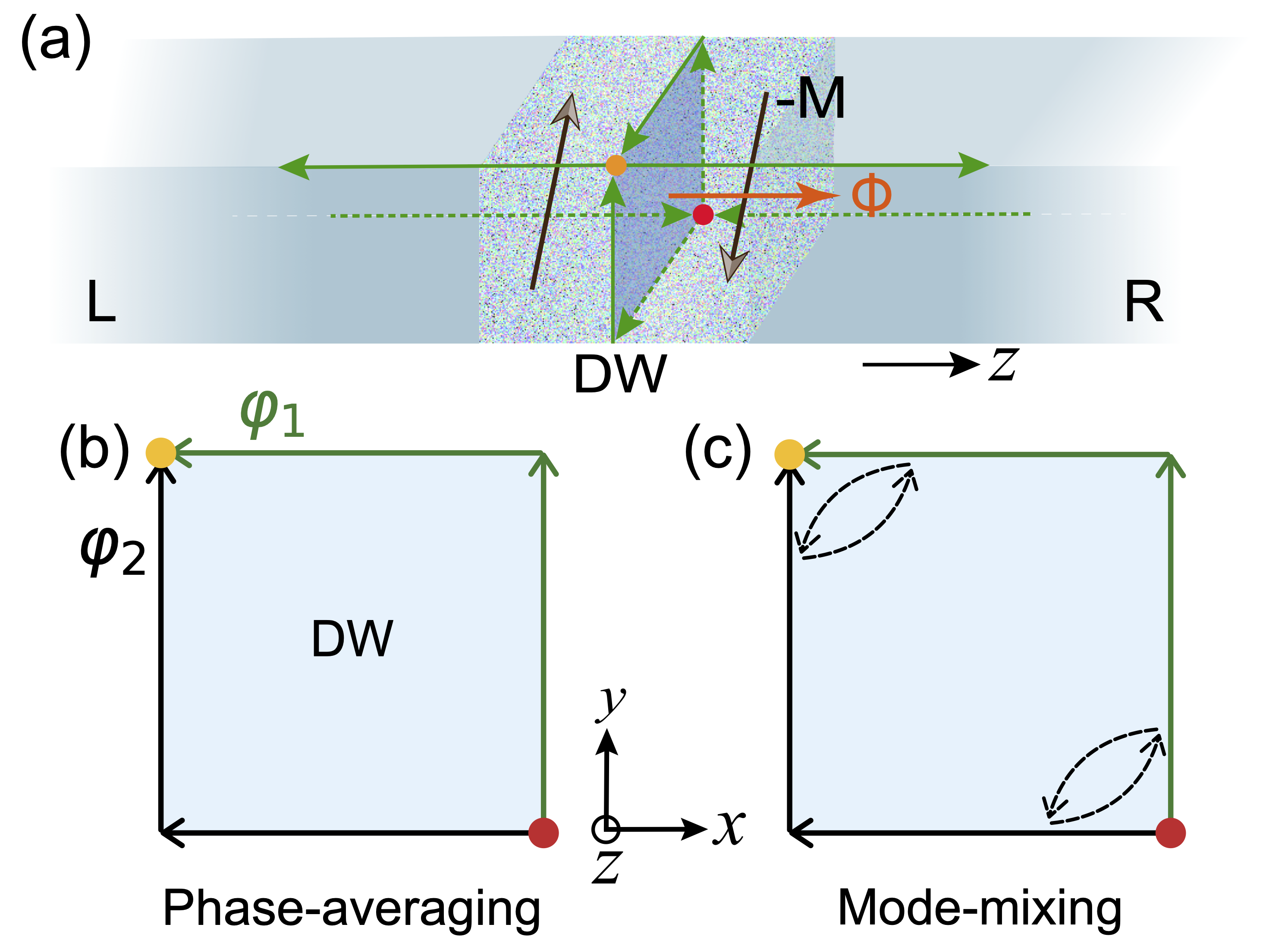}
    \caption{\label{fig_device}
(a) Cartoon illustration of the AB interferometer based on a 3D SOTI nanowire containing a magnetic domain wall (DW) structure, where a magnetic flux $\Phi$ (indicated by the orange arrow) is threaded through the DW to tune the two-terminal conductance. Green arrows indicate the propagation of the THSs and TESs. On-site disorder is applied to the central scattering region encompassing the DW (shadowed area). (b) and (c): Schematic diagrams for the PAR at weak disorder and MMR at strong disorder in the magnetic DW. The black and green arrows denote the two co-propagating TESs. In (b), electrons traveling along the upper-right and lower-left arms acquire distinct dynamical phases $\varphi_1$ and $\varphi_2$ due to the presence of electrostatic disorder. In (c), the dashed curves represent the mixing between the upper-right and lower-left TESs.
}
\end{figure}

In this paper, we investigate electronic transport across a magnetic DW in a 3D SOTI~\cite{Schindler2018HigherA, Schindler2018HigherB, Noguchi2021Evidence, Geier2018Second, Langbehn2017Reflection, Hou2023Realization, Hou2025Magnetic} subject to Anderson disorder, and propose it as an ideal platform for concurrently studying both the phase-averaging and mode-mixing phenomena. As illustrated in Fig.~\ref{fig_device}(a), by magnetically dopping a 3D TI~\cite{Hor2010Development, Xu2012Hedgehog, Rienks2019Large} along the diagonal direction~\cite{Langbehn2017Reflection, Hou2023Realization, Hou2025Magnetic}, two chiral one-dimensional (1D) topological hinge states (THSs) appear on opposite hinges. Constructing a magnetic DW further induces four 1D topological edge states (TESs) surrounding the DW, as a result of the sign-reversal of the out-of-plane magnetization across the DW. These TESs link the THSs into a closed loop, forming a perfect AB interferometer with a conductance-magnetic flux relation: $G = (0.5 - 0.5 \cos{\Phi} ) e^2/h$~\cite{Hou2025Magnetic}. As disorder is introduced, the AB oscillation is suppressed, while a half-quantized CP of $0.5 e^2/h$ emerges, alongside a conductance-fluctuation ($\sigma_G$) plateau of $\sim0.35 e^2/h$ [see Fig.~\ref{fig_mainresult}(c,d) the main results]. By developing a theoretical framework accounting for the random-phase interference of the two co-propagating TESs at moderate disorder [see Fig.~\ref{fig_device}(b) the green and black arrows], we identify this $\sigma_G$ plateau as a signature of the PAR. As disorder strength increases further, the half-quantized CP persists, whereas the $\sigma_G$ plateau collapses and then reaches a second plateau of $\sim 0.29 e^2/h$. This second $\sigma_G$ plateau is a hallmark of the MMR, arising from the complete inter-mode mixing of the TESs driven by strong disorder [see Fig.~\ref{fig_device}(c)]. The evolution of these $\sigma_G$ plateaus reveals a clear disorder-induced crossover from PAR to MMR. 

Furthermore, our theory of PAR predicts a U-shaped beta-distribution for the conductance, whereas the distribution becomes uniform in the MMR. These are validated by large-scale numerical simulations. To gain direct insight into the evolution of these transport regimes, we plot the spatial current density distribution at the DW across various disorder strengths. At weak or moderate disorder, the local current remains unidirectional and exhibits a ``hollow" spatial profile, signifying that the AB interferometer remains intact. In contrast, under strong disorder, this unidirectional propagation is disrupted, and the current density instead forms a ``diffusive cloud" profile, providing direct evidence of mode-mixing. Finally, we calculate the Fano factor $F$ associated with shot-noise measurements. This factor exhibits a two-step plateau with values of 1/4 and 1/3 in the PAR and MMR, respectively, offering a clear metric for experimental verification. Our work highlights second-order cumulants of conductance—specifically fluctuations and the Fano factor—as essential fingerprints for identifying transport regimes where the half-quantized CP alone is insufficient. These findings offer a promising avenue for the design and control of DW-based electronics with the aid of disorder-engineering.

The remainder of this paper is organized as follows: Section~\ref{sec_model_methods} introduces the tight-binding model of the 3D SOTI with a magnetic DW, and the method for quantum transport. Section~\ref{sec_main_results} presents the main numerical results. Section~\ref{sec_phase_averaging} develops the theory on the PAR and provides numerical verifications including the conductance statistics and local current density distribution. Section~\ref{sec_mode_mixing} details the theory of the MMR and discusses the evolution of transport regimes. Section~\ref{sec_Fano} presents the results on the Fano factor. Section~\ref{sec_disorder_location} explores how varying the spatial distribution of disorder affects transport properties. Finally, Sec.~\ref{sec_conclusion} provides a summary and concluding remarks.

\section{MODEL AND METHODS}
\label{sec_model_methods}
\subsection{Tight-binding Model for the 3D SOTI with a DW}
We consider a 3D SOTI nanowire embedded with a magnetic DW as shown in Fig.~\ref{fig_device},  where Anderson disorder is applied around the DW. A uniform magnetic field ${\bf B}$ is applied along the nanowire direction ($z$-axis) to induce a magnetic flux through the DW. To describe this system, we adopt a four-band tight-binding Hamiltonian on a cubic lattice~\cite{Hosur2010Chiral, Hosur2011Majorana, Shen2013Topological}:
\begin{equation}
\label{eq_H}
\hat{H} = \sum_{\mathbf{i}} c_{\mathbf{i}}^\dagger \mathcal{E}_{{\bf i}} c_{\mathbf{i}} + \sum_{\mathbf{i}, \alpha=x,y,z} \left( c_{\mathbf{i}}^\dagger t_{\alpha} e^{i \phi_{{\bf i}, {\bf i} + {\bf e}_\alpha}} c_{\mathbf{i}+ {\bf e}_\alpha} + \text{H.c.} \right),
\end{equation}
where $c_{\mathbf{i}}^\dagger = [c^{\dagger}_{1{\uparrow}} ({\bf i}), c^{\dagger}_{ 1{\downarrow}} ({\bf i}), c^{\dagger}_{2{\uparrow}} ({\bf i}),  c^{\dagger}_{ 2{\downarrow}} ({\bf i}) ] $ is the four-component creation operator at site $\mathbf{i}$, with the subscripts 1(2)  and $\uparrow (\downarrow)$ denoting the electron's orbital and spin, respectively. The on-site term $\mathcal{E}_{\bf i}$ reads:
\begin{equation}
\label{eq_onsite}
\mathcal{E}_{\bf i} = m_0 \sigma_z \otimes s_0 + \sigma_0 \otimes {\bf M}_{\bf i} \cdot {\bf s} + U_{\bf i} \sigma_0 \otimes s_0,
\end{equation} 
where the Pauli matrices $\sigma_{ \alpha}$ and $s_{ \alpha}$ act on orbital and spin spaces, respectively. Here the first term in Eq.~(\ref{eq_onsite}) describes the typical on-site mass term for a pristine 3D TI. The second term describes an effective exchange field originating from the magnetic doping with Mn or Cr atoms onto 3D TI~\cite{Hor2010Development, Xu2012Hedgehog, Rienks2019Large}, which is essential for promoting the 3D TI into a SOTI by setting the magnetization orientation along the diagonal direction of the $x-y$ plane: $\pm {\bf n}_{110} = \pm  (1/\sqrt{2}, 1/\sqrt{2}, 0)$. Under diagonal magnetization, a gap opens on the four facets of the 3D TI nanowire due to the nonzero net out-of-plane magnetization. Furthermore, the surface gap reverses its sign on two opposite hinges of the nanowire, generating a pair of 1D THSs [see Fig.~\ref{fig_device}(a)]. The third term accounts for Anderson-type disorder within the central scattering region of length $L$. Unless otherwise specified, we assume this disorder is present throughout the entire central region. The disorder potential $U_{\mathbf{i}}$ is uniformly distributed within $ [-W/2, W/2]$ for $-L/2 <  z_{\bf i} \leq L/2$ and vanishes elsewhere, where $W$ characterizes the disorder strength.

The nearest-neighbor hopping matrix $t_{\alpha}$ takes the specific form:
\begin{equation}
 t_{\alpha} = \frac{-i \hbar v_F}{2a} \sigma_x \otimes s_\alpha - \frac{m_1}{2} \sigma_z \otimes s_0.
\end{equation}
Here, $v_F$ represents the Fermi velocity, $a$ is the lattice constant, and $m_1$ is the mass term which is essential for  the band inversion and determines the topological phase. To incorporate the effect of magnetic field, we add a phase $\phi_{{\bf i}, {\bf i} + {\bf e}_{\alpha}} = \frac{e}{\hbar} \int_{\mathbf{r}_{\mathbf{i}}}^{{\mathbf{r}_{\mathbf{i}}} + {\bf e}_{\alpha}} {\bf A} \cdot d {\bf r}$ into the hopping matrix $t_\alpha$ via the Peierls substitution, where a Landau gauge of ${\bf A} = (-By, 0, 0)$ has been chosen.

To construct a magnetic DW, we set the magnetization vector as:
\begin{equation}
{\bf M}_{\bf i} =
\begin{cases}
 M {\bf n}_{110}, & \text{for } z_{\bf i} < 0, \\
-M {\bf n}_{110}, & \text{for } z_{\bf i} \geq 0,
\end{cases}
\end{equation}
where $M>0$ characterizes the magnetization strength. In this case, another sign reversal of the surface gap across the DW happens for all the four facets, generating four additional 1D TESs on the edges of the DW [see the four arrows surrounding the DW in Fig.~\ref{fig_device}(a)]. Those states connect the counter-propagating THSs on both sides of the nanowire to form an AB interferometer, allowing a magnetic-flux-tunable way on the transmission across the DW~\cite{Hou2025Magnetic}.

\subsection{Methods for quantum transport}
The nanowire in Fig.~\ref{fig_device}(a) can also be regarded as a two-terminal quantum transport device. It consists of a central scattering region containing the disordered magnetic DW, coupled to the left and right electrodes composed of clean and semi-infinite 3D SOTI nanowires. To calculate the conductance, we employ the nonequilibrium Green's function (NEGF) method. First, we calculate the surface Green's functions ${\bf g}^r_{L(R)}(E_F)$ of the left (right) electrodes using the recursive Green's function method \cite{Surface_GF}, where $E_F$ is the Fermi energy. The self-energy coupled to the central region is calculated as: ${\bm \Sigma}^r_{L(R)}={\bf H}_{cL(R)}{\bf g}^r_{L(R)}(E_F) {\bf H}^\dagger_{cL(R)}$ with ${\bf H}_{cL(R)}$ the coupling matrix from the central region to the left(right) electrode. The retarded Green's function of the central region is defined as ${\bf G}^r_c(E_F)=\left[ (E_F+i 0^+){\bf I}-{\bf H}_c-{\bm \Sigma}^r_L-{\bm \Sigma}^r_R \right]^{-1}$, with ${\bf H}_c$ the Hamiltonian matrix of the central region. The transmission coefficient through the central scattering region is calculated as~\cite{Meir1992LandauerFormula, Jauho1994TransportResonant, Datta1995Mesoscopic}:
\begin{eqnarray}
T (E_F) =  {\rm Tr}\left[ {\bm \Gamma}_L {\bf G}^r_c  {\bm \Gamma}_R {\bf G}^a_c  \right],
\end{eqnarray}
where ${\bm \Gamma}_{L(R)}\equiv i \left[ {\bm \Sigma}^r_{L(R)}-({\bm \Sigma}^r_{L(R)})^\dagger \right]$ is the linewidth function for electrode L(R) and ${\bf G}^a_c =\left( {\bf G}^r_c \right) ^\dagger$ is the advanced Green's function. The zero-temperature differential conductance is $G = \frac{e^2}{h} T (E_F)$ according to the Landauer-B{\"u}ttiker formula~\cite{Buttiker1985Generalized}. For convenience, the conductance unit $e^2/h$ will be omitted hereafter. 

In Sections~\ref{sec_phase_averaging} and \ref{sec_mode_mixing} we plot the spatial current density distribution to visualize the evolution of the transport regimes. To this end, we define the nonequilibrium local current vector at site ${\bf i}$ as ${\bf J}_{\bf i} = (J_{\bf i}^x, J_{\bf i}^y, J_{\bf i}^z )$, with the component given by
\begin{align}
J_{\bf i}^{\alpha} = (J_{{\bf i} \rightarrow {\bf i}+ {\bf e}_\alpha} + J_{{\bf i} - {\bf e}_\alpha  \to  {\bf i}} ) /2, \ \alpha = x, y, z. 
\end{align} 
Here $J_{{\bf i}\to {\bf j}}$ is the amplitude of the local bond current between two nearest-neighbor sites ${\bf i}$ and ${\bf j}$, expressed in terms of the Green's function as follows~\cite{{Nikoli2006Imaging, Zarbo2007Spatial, Jiang2009Numerical}}:
\begin{align}
J_{{\bf i} \to {\bf j}} = \frac{2e^2}{h}  {\rm Im} \{ {\rm Tr} [ H_{\bf i, j} G^n_{\bf j, i} (E_F) ] \} (V_L - V_R),
\end{align}
where $H_{\bf i, j}$ is the hopping matrix between sites $\bf i$ and $\bf j$, $V_{L(R)}$ is the voltage at electrode L(R), and $G^n(E_F) = G^r(E_F) \Gamma_L(E_F) G^a(E_F)$.    

In the numerical calculations, we set the Fermi velocity $v_F = 5 \times 10^5$ m/s and the lattice constant $a=2.2$ nm to align our model with realistic materials such as Bi$_2$Se$_3$ or Bi$_2$Te$_3$~\cite{Zhang2009Topological, Liu2010Model, Qi2011Topological, Xia2009Observation, Chen2009Experimental}. We define $E_0 = \hbar v_F /a \approx 150$ meV as the energy unit. By setting $m_0 = 2E_0$ and $m_1=E_0$, the system enters the strong TI phase; the magnetization term in Eq.~(\ref{eq_onsite}) further drives it into a SOTI. For the nanowire cross-section in the $x-y$ plane, we employ a discretization of $N \times N$ lattice sites under open boundary conditions. The reduced magnetic flux through the nanowire is given by $\Phi = (N-1)^2 B a^2 \pi /\Phi_0 $, and the length of the disordered central region is fixed at $L$ sites. Throughout this work, we set $N=20$ and $L=10$. The magnetization strength is chosen as $M=0.5 E_0$, which opens a surface gap of approximately $75$ meV. At this strength, the localization length of the THSs along the surface is as small as $2.9 a$ (following the calculation method in Ref.~\cite{Hou2025Magnetic}), effectively eliminating coupling between THSs on opposite hinges. While experimental sample sizes are typically much larger than our model, the longer localization lengths associated with smaller surface gaps would still not induce inter-hinge coupling in such systems. Thus, our model remains a reliable simulation of realistic experimental devices.

\section{Main Numerical Results}
\label{sec_main_results}
% === Figure 2 ===
\begin{figure}[t]
  \centering
  \includegraphics[width=0.48\textwidth]{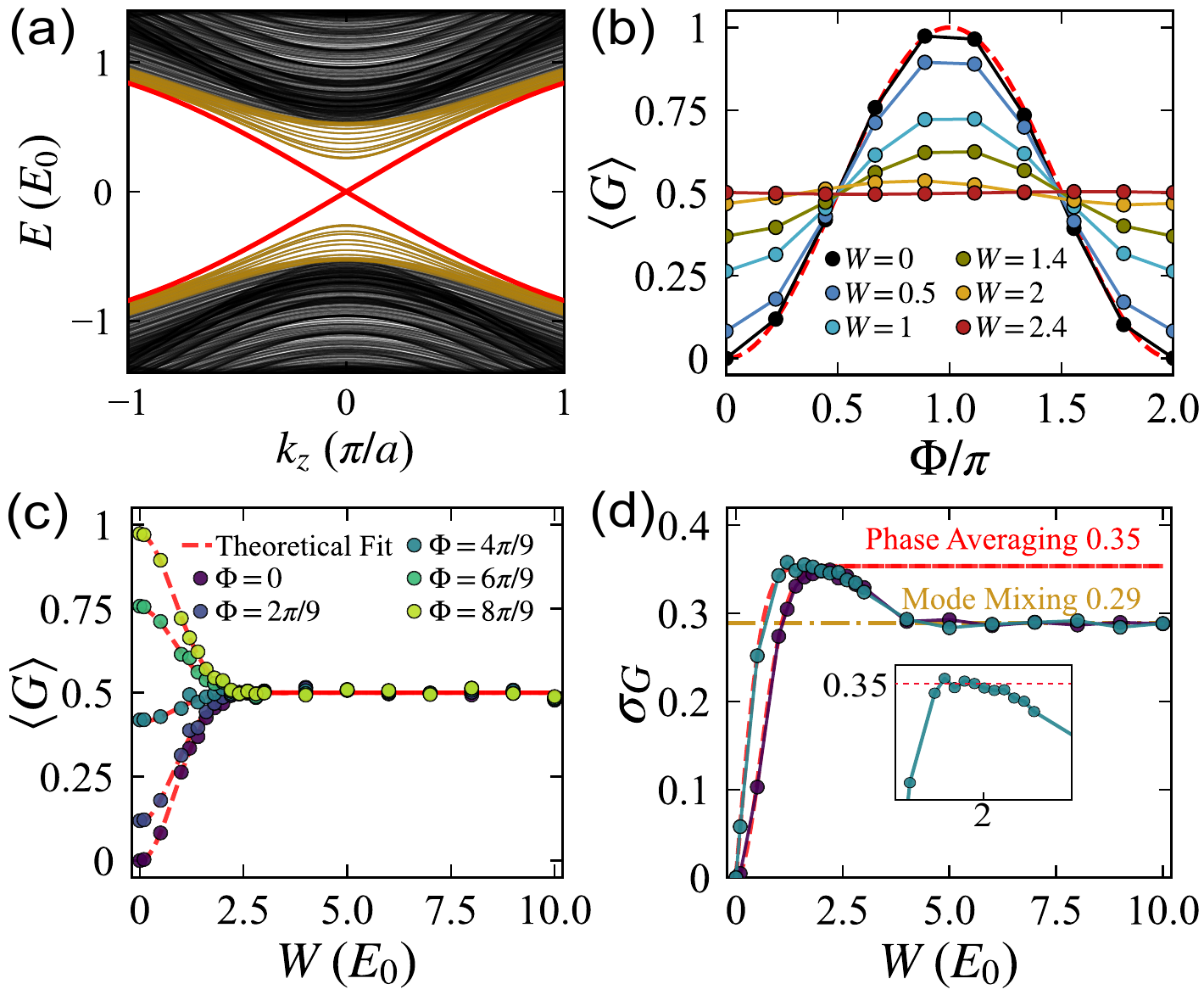} 
  \caption{\label{fig_mainresult}
  Numerical results for quantum transport across the magnetic DW in a 3D SOTI. (a) Energy band structure of a uniform SOTI nanowire with magnetization vector ${\bf M} =(\sqrt{2}/4, \sqrt{2}/4, 0) E_0 $. The 1D THSs, 2D surface states, and 3D bulk states are shown in red, dark yellow, and black curves, respectively. (b) Ensemble-averaged conductance $\langle G \rangle$ as a function of the magnetic flux $\Phi$ under different disorder strengths $W$. The red dashed curve is the analytical fitting using the formula $G(\Phi)= 0.5 - 0.5 \cos{\Phi}$. (c) and (d) Average conductance $\langle G \rangle$ and conductance fluctuation $\sigma_G$ versus disorder strength $W$ at different magnetic fluxes $\Phi$. The red dashed curves in (c) and (d) are theoretical fits for $\langle G \rangle$ and $\sigma_G$, using analytical formulas in Eq.~(\ref{eq_Gdis_analy}) and Eq.~(\ref{eq_sigmaG_analy}), respectively. The fitting parameter $\chi \approx 0.733$. The inset in (d) shows a magnified view of the fluctuation plateau. For clarity, only two values of $\Phi$ (0 and $4\pi/9$) are shown in (d). For (b)--(d), the Fermi energy is $E_F = 0.02 E_0$, and the results are averaged over 1200 disorder configurations.
  }
\end{figure}

In this section we present the the main transport results across the disordered DW. In Fig.~\ref{fig_mainresult}(a) we first plot the band structure of a pristine SOTI nanowire with magnetization along ${\bf n}_{110}$ direction. Here the linearly dispersing chiral THSs crossing the band gap are shown in red. The THS with negative(positive) slopes is indicated by the solid (dashed) arrows in the left terminal of Fig.~\ref{fig_device}(a). A surface gap $\Delta_{\rm surf} \approx 75$ meV is opened by the diagonal magnetization. Above this gap, the two-dimensional (2D) surface states and 3D bulk states are identified by their wavefunction distributions and are depicted as yellow and black curves, respectively. For electron energies within the surface gap, quantum transport in the electrodes is mediated by these counter-propagating chiral THSs. At the magnetic DW, the THSs are connected by 1D TESs [see the four arrows on the DW edges in Fig.~\ref{fig_device}(a)]. Together, these states form a perfect AB interferometer, with the conductance governed by the relation $G(\Phi)= 0.5 - 0.5 \cos{\Phi} $. 

Figure~\ref{fig_mainresult}(b) shows the conductance as a function of the magnetic flux $\Phi$ for various disorder strengths. In the clean limit ($W=0$), the conductance exhibits AB oscillations, in perfect agreement with the sinusoidal relation $G(\Phi)$ (see the red dashed fitting curve). At weak disorder (e.g. $W=0.1 E_0$), the ensemble-averaged conductance $\langle G \rangle$ retains its oscillatory pattern, though the oscillation amplitude is noticeably suppressed. As $W$ is increased further, the oscillation amplitude decays to nearly zero for $W\geq 2 E_0$; in this regime, the conductance becomes independent of $\Phi$ and settles into a half-quantized plateau at 0.5. 

Figure~\ref{fig_mainresult}(c) illustrates the $W$-dependence of $\langle G \rangle $ at different flux values. Note that here we restrict $\Phi$ to the range $[0,  \pi]$, as $\langle G \rangle$ is nearly symmetric about $\Phi=\pi$~\footnote{The discrepancy between the numerical conductance and the analytical formula stems from the percolation of the TESs into the bulk, which makes the true magnetic flux enclosed by the enclosed loop smaller than $\Phi$. The nonsymmetric relation of $\langle G \rangle (\Phi)$ about $\Phi=\pi$ stems from the same reason.}, as observed in Fig.~\ref{fig_mainresult}(b). This can be understood from the fact that, in the clean limit, the system at $-\Phi$ can be mapped to the system at $\Phi$ via a mirror transformation about the (110) plane~\cite{Hou2025Magnetic}. While the conductance varies with $\Phi$ at weak disorder, the values converge to the half-quantized plateau once $W> 2E_0$. This 0.5-CP persists over a wide range of disorder strengths, remaining robust even in the strong disorder limit ($ W=10E_0$).

To examine the statistical properties of the conductance $G$, we define the conductance fluctuation as $\sigma_G \equiv \sqrt{ \langle G^2 \rangle - \langle G \rangle ^2}$ and plot its $W$-dependence in Fig.~\ref{fig_mainresult}(d). As $W$ increases, $\sigma_G$ increases monotonically until it saturates to a plateau of $\sim 0.35$ near $W=2 E_0$. Unlike the $\langle G \rangle$-plateau, the initial $\sigma_G$ plateau collapses as $W$ increases further. Eventually, $\sigma_G$ reaches a second plateau of $\sim 0.29$ for $W>4E_0$. This latter plateau is more stable than the first, persisting even in the strong disorder limit.

\section{Theory of Phase-averaging}
\label{sec_phase_averaging}
The appearance of two distinct $\sigma_G$ plateaus despite an identical half-quantized $\langle G \rangle$ plateau indicates the presence of two underlying transport mechanisms. Since the first $\sigma_G$ plateau emerges at moderate disorder, the 1D THSs and TESs is assumed to keep their unidirectional propagation, and the primary effect of disorder is the introduction of random dynamical phases along the propagation paths. Thus, one is tempted to attribute the first $\sigma_G$ plateau to a phase-averaging mechanism. 

Before providing a detailed derivation, we first present a global picture of the PAR, which provides a simple and straightforward explanation for the emergence of $\langle G \rangle$ and $\sigma_G$ plateaus. In the clean limit, the conductance follows a perfect sinusoidal relation on $\Phi$. Once disorder is introduced, extra dynamical phases $\varphi_{1(2)}$ are acquired along the paths of the upper-right and lower-left TES arms [see the green and black arrows in Fig.~\ref{fig_device}(b)]. This results in a dynamical phase difference $\Delta \varphi \equiv  \varphi_2 - \varphi_1$ which, together with the magnetic flux, constitutes an effective flux $\Phi_{\rm eff} \equiv \Phi + \Delta \varphi $. The conductance is then governed by the relation
\begin{align}
\label{eq_G_Phieff}
G(\Phi_{\rm eff}) = 0.5- 0.5 \cos(\Phi_{\rm eff}) .
\end{align} 
When the path is long, or the disorder strength is relatively strong, the effective flux $\Phi_{\rm eff}$ becomes fully randomized and uniformly distributed within $[0, 2\pi]$, with the probability density function
\begin{align}
\label{eq_P_Phi}
P_{\Phi} = {1}/{2\pi}.
\end{align}
Then the ensemble-averaged conductance is calculated as:
\begin{align}
\label{eq_G_phase}
\langle G \rangle = \int_{0}^{2\pi} G(\Phi_{\rm eff}) P_{\Phi} \, {\rm d}\Phi_{\rm eff} = {1}/{2} .
\end{align}
The variance of conductance is
\begin{align}
\Delta G & = \langle(G - \langle G \rangle )^{2}\rangle = \int_{0}^{2\pi}[G(\Phi_{\rm eff})- {1}/{2} ]^{2} P_{\Phi} {\rm d} \Phi_{\rm eff}  \\ \notag 
&={1}/{8}	 
\end{align}
This yields the conductance fluctuation 
\begin{align}
\label{eq_sigmaG_phase}
\sigma_G = \sqrt{ \Delta G } = \frac{\sqrt{2}}{4}  \approx 0.354 .
\end{align}
The results in Eq.~(\ref{eq_G_phase}) and Eq.~(\ref{eq_sigmaG_phase}) are in perfect agreement with Fig.~\ref{fig_mainresult}(c, d) at $W$ around $2 E_0$.

\subsection{Detailed derivation on the expressions of conductance and fluctuation}

In this subsection, we provide a detailed derivation of the analytical expressions for the disorder-strength ($W$) dependence of $\langle G \rangle$ and $\sigma_G$. This theoretical framework is applicable to both the weak and moderate disorder regimes.

For the co-propagating TESs at the DW edges, the kinetic energy is determined by the linear dispersion relation $E_{\text{ki}} = \hbar v_{F} k$. Here the wavevector $k$ is always aligned with the propagation direction of the TESs. 

The dynamical phase accumulated by the TESs along path $C_{n}$ is given by the following line integral:
\begin{align}
\varphi_n = \int_{C_n} k(l)\, dl  .
\end{align}
Here $n=1(2)$ denotes the upper-right (lower-left) arm. Within our tight-binding model with lattice constant $a$, the paths consist of discrete sites indexed by $j$. The wavevector at site $j$ is given by
\begin{align}
k_{j} = \frac{E_{\rm ki}}{\hbar v_{F}}
   = \frac{E - U_{j}}{\hbar v_{F}} ,
\end{align}
where $E$ is the total energy and $U_j$ is the local electrostatic potential induced by disorder. Consequently, the continuous dynamical phase is replaced by the discrete summation
\begin{align}
\varphi_{n}&=\sum_{j \in C_{n}}k_{j} a  =\sum_{j \in C_{n}}(\widetilde{E}-\widetilde{U}_{j}) ,
\end{align}
where we have defined the dimensionless energy $\widetilde{E} \equiv E/E_0$, and the dimensionless on-site disorder potential $\widetilde{U}_j \equiv U_j/E_0$. The latter is uniformly distributed within $[-\widetilde{W}/{2}, \widetilde{W}/{2}]$. Note that $\widetilde{W}$ does not simply represent the disorder strength $W$ normalized by the energy unit; rather it is defined as
\begin{align}
\widetilde{W} \equiv \chi {W} / {E_0},
\end{align}
where $\chi$ is a modification factor that accounts for the spatial distribution of the TESs. While our derivation treats the TESs as idealized 1D channels, the actual disorder potential experienced by these states deviates from the bare on-site potential due to their finite broadening at the DW interface. This effective reduction in the perceived disorder strength is captured by the factor $\chi$, whose value is determined via numerical fitting. 

The dynamical phase difference between path $C_2$ and $C_1$ is:
\begin{align}
\label{eq_DeltaPhi}
\Delta\varphi & = \varphi_{2}-\varphi_{1} = \sum_{j \in C_{2}}\left(\widetilde{E}-\widetilde{U}_{j}\right)-\sum_{j  \in C_{1}}\left(\widetilde{E}-\widetilde{U}_{j}\right) \\ \notag
&=\sum_{j \in C_{2}}	(-\widetilde{U}_{j} )+	\sum_{j \in C_{1}} \widetilde{U}_{j}
=\sum_{j\in C_{1}\cup C_{2}}\widetilde{U}_{j} .
\end{align}
Here we have used the square geometry of the DW, where the path lengths are approximately equal, $N_{C_1} = N_{C_2} \approx 2N$. The minus sign in the second line is eliminated by redefining the potential at sites $j \in C_2$ as $\widetilde{U}_{j} \rightarrow  - \widetilde{U}_{j}$, which is also uniformly distributed within $[-\widetilde{W}/{2}, \widetilde{W}/{2}]$. 

Given that the disorder potentials $\widetilde{U}_j$ are spatially uncorrelated, the dynamical phase difference $\Delta \varphi$ in Eq.~(\ref{eq_DeltaPhi}) can be treated as a sum of $4N$ independent random variables. For sufficiently large $N$, the central limit theorem ensures that $\Delta \varphi$ follows a normal distribution:
\begin{align}
\label{eq_Pvarphi}
P_{\varphi}(\Delta \varphi) = 
\frac{1}{\sqrt{2\pi\sigma_{\varphi}^{2}}} e^{-\frac{(\Delta\varphi)^{2}}{2 \sigma_{\varphi}^{2}}} ,
\end{align}
where the mean value is set to zero by definition: $\langle \Delta \varphi \rangle = \sum_{j\in C_{1}\cup C_{2}} \langle  \widetilde{U}_{j} \rangle = 0$, and $\sigma_{ \varphi}$ denotes the standard deviation (std) of $\Delta \varphi$, which follows
\begin{align}
\sigma_{\varphi} = \sqrt{4 N } \sigma_U = \frac{\sqrt{3N}}{3}\widetilde{W}
\end{align}
where we have utilized the std of the uniform distribution for $\widetilde{U}_j$, given by $\sigma_U =  \sqrt{3} \widetilde{W}/6$.  

The conductance across the DW for a specific disorder configuration is given by $G_{\rm dis} (\Phi) = G(\Phi + \Delta \varphi) = 0.5 - 0.5 \cos{ (\Phi + \Delta \varphi)}$, which is $2\pi$-periodic with respect to $\Delta \varphi$. Consequently, the effective distribution of $\Delta \varphi$ can be obtained by folding the normal distribution into the interval $[0, 2\pi]$:
\begin{align}
\label{eq_Pvarphi2pi}
P_{\varphi} ^ { [0, 2\pi]} (\Delta \varphi)
&=\sum_{n = -\infty}^{+\infty} P_{\varphi} (\Delta\varphi+2\pi n)\\  \notag
&=  \frac{1}{\sqrt{2\pi \sigma_{\varphi}^{2}}}  \sum_{n = -\infty}^{+\infty} \exp[-(\Delta\varphi+2\pi n)^{2}/2 \sigma_{\varphi}^{2} ] \\ \notag
&= \frac{1}{2\pi} \theta_3 \left( \frac{\Delta \varphi}{2}, e^{-\sigma_\varphi^2/2} \right)	, \quad 0\leq \Delta \varphi \leq 2 \pi .\notag
\end{align}
Here $n$ is an integer and $\theta_3$ denotes the Jacobi theta function. In the final step we have utilized the modular identity: $\theta_3 \left( \frac{z}{\tau}, e^{-i \pi/\tau} \right) = (-i \tau)^{1/2} \exp{\left( \frac{i z^2}{\pi \tau} \right) } \theta_3(z, q)$ with $q = e^{i \pi \tau}$.

Next, we calculate the ensemble-averaged conductance $\langle G \rangle$. Given the $2\pi$-periodicity of the sinusoidal conductance function, it is more computationally convenient to employ the original Gaussian distribution $P_{\varphi}$ from Eq.~(\ref{eq_Pvarphi}) rather than the folded distribution $P_{\varphi} ^ { [0, 2\pi]}$ from Eq.~(\ref{eq_Pvarphi2pi}). The resulting ensemble average is (see Appendix~\ref{app_Gaussian_integral} for the detailed algebra):
\begin{align}
\label{eq_Gdis_analy}
\langle G_{\rm dis}(\Phi)\rangle
&=\int_{-\infty}^{+\infty} G (\Phi + \Delta \varphi)\, P_{\varphi}(\Delta \varphi)\, d\Delta \varphi  \\ \notag
&= \frac{1}{2}-\frac{1}{2}e^{-\frac{N}{6}\widetilde{W}^2}\cos\Phi .   \notag
\end{align}

The variance of the conductance is
\begin{align}
\label{eq_VarG_analy}
{\rm Var} (G) =\int_{-\infty}^{+\infty}\left[G(\Phi  + \Delta \varphi )-\langle G_{\rm dis}(\Phi) \rangle \right]^{2}P_{\varphi}( \Delta\varphi )d\Delta\varphi .
\end{align}
Following Appendix~\ref{app_Gaussian_integral}, we obtain the analytical expression for the conductance fluctuation:
\begin{align}
\label{eq_sigmaG_analy}
\sigma_{G} &= \sqrt{{\rm Var} (G) } \\ \notag
&= \sqrt{ \frac{1}{8} \left[ 1- \cos(2\Phi)e^{-\frac{N}{3}\widetilde{W}^{2}} \right] ( 1- e^{-\frac{N}{3}\widetilde{W}^{2}} ) } .
\end{align}

Next, we examine the behavior of the system in two distinct limits of $N \widetilde{W}^2 $:  \\
(1) The Clean Limit ($N \widetilde{W}^2 \rightarrow 0$) \\
In the weak-disorder regime, the ensemble-averaged conductance can be approximated as
\begin{align}
\langle G_{\rm dis}(\Phi)\rangle =
G(\Phi)+\frac{N\cos\Phi}{12} \widetilde{W}^{2}	,
\end{align}
showing a parabolic dependence on $\widetilde{W}$. For $\Phi$ not in the immediate vicinity of zero or $\pi$, the conductance fluctuation is linear to $\widetilde{W}$:
\begin{align}
\sigma_G = \sin{\Phi} \frac{\sqrt{3N}}{6}  \widetilde{W} 	,
\end{align}
which vanishes as $\widetilde{W} \rightarrow 0$, as expected. By performing a first-order expansion, $G_{\rm dis} (\Phi) \approx   G(\Phi)+  \frac{\sin{\Phi}}{2}  \Delta \varphi$, we find that $G_{\rm dis}$ follows a normal distribution:
\begin{align}
\label{eq_PG_Gauss}
P_G(G_{\rm dis}) = \frac{1}{\sqrt{2\pi \sigma_G ^2}} \exp{\left( -{ [G_{\rm dis}-G(\Phi)]^2} /{2 \sigma^2_G} \right) } .
\end{align}
(2) The Phase-Averaging Limit ($N \widetilde{W}^2 \rightarrow \infty$) 	\\
In the ``dirty" or moderate-disorder case, where the phase uncertainty becomes large, we obtain:
\begin{align}
\langle G_{\rm dis} (\Phi)\rangle = \frac{1}{2}, \quad \sigma_G =  \frac{\sqrt{2}}4 .
\end{align}
These values correspond precisely to the first $\sigma_G$-plateau observed at moderate disorder in Fig.~\ref{fig_mainresult}(d), marking the onset of the PAR. 

Besides, we get $\sigma_{\varphi} \rightarrow + \infty$ in this limit, and the Jacobi theta function in Eq.~(\ref{eq_Pvarphi2pi}) approaches unity, $ \theta_3 \left( \frac{\Delta \varphi}{2}, e^{-\sigma_\varphi^2/2} \right) \rightarrow 1$. This yields a uniform distribution function $P_{\varphi} ^ { [0, 2\pi]} (\Delta \varphi) = 1/2\pi$, consistent with the phase-averaging result summarized in Eq.~(\ref{eq_P_Phi}). 

\begin{figure*}[t!]
  \centering
  \includegraphics[width=0.95\textwidth]{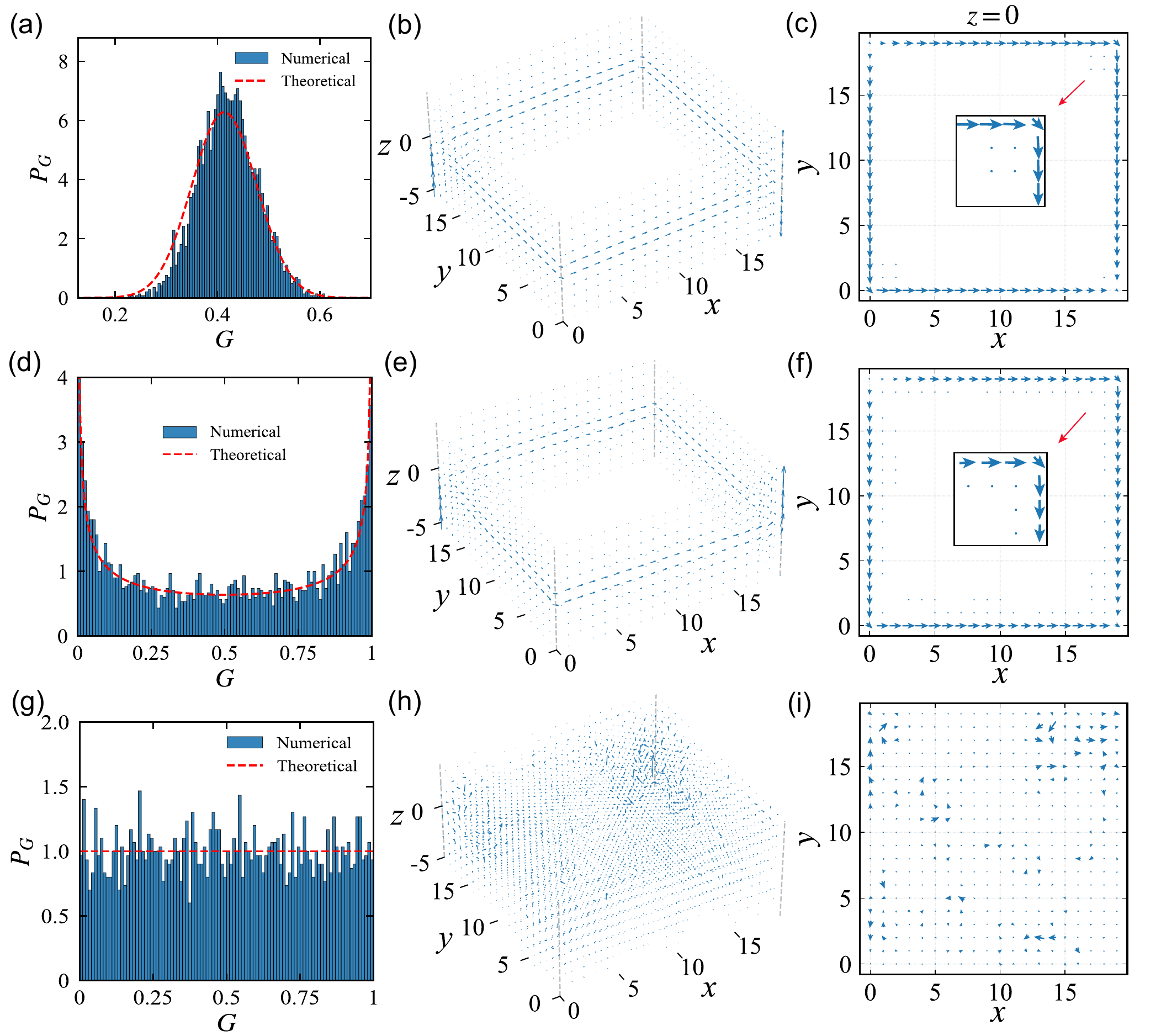} 
  \caption{\label{fig_currentdensity}
Statistics of conductance $G$ and current density distribution at different disorder strengths: (a)--(c) Weak disorder ($W=0.1 E_0$); (d)--(f) Moderate disorder ($W=2.2 E_0$) at which the PAR occurs; (g)--(i) Strong disorder ($W=9.0 E_0$) at which the MMR occurs. In (a, d, g) the numerically calculated probability distribution $P_G$ are present, which exhibit a Gaussian profile, a U-shaped profile, and a flat uniform profile, respectively. These results are in excellent agreement with theoretical predictions shown by red dashed curves. (b, e, h): The 3D visualization of the current density; (c, f, i): Cross-sectional current density plot at $z=0$. For all panels, the magnetic flux is fixed at $\Phi = 4\pi/9$, the Fermi energy is $E_F = 0.02 E_0$, and the statistical distributions of $P_G$ are obtained from an ensemble of 3000 disorder configurations with a bin-width of $h=0.01$. The insets in (c) and (f) provide magnified views of the highly localized hinge currents at the top-right corner, as indicated by the red arrows.
}
\end{figure*}

As the phase difference $\Delta \varphi$ becomes uniformly distributed within $[0, 2\pi]$, the probability distribution function of the conductance $G_{\rm dis}$ can be derived as:
\begin{align}
\label{eq_PG_U}
P_G(G_{\rm dis}) &= 2 P_{\varphi}^{[0, 2\pi]} [\Delta \varphi( G_{\rm dis})] \left| \frac{{\rm d} \Delta \varphi ( G_{\rm dis}) }{{\rm d }G_{\rm dis}} \right| \\ \notag
&= \frac{1}{\pi \sqrt{G_{\rm dis}(1-G_{\rm dis})}},
\end{align}
where $\Delta \varphi(G_{\rm dis}) = \arccos(1-2G_{\rm dis}) - \Phi$ represents the inverse relationship between the phase difference and the conductance. The factor of $2$ in the first line accounts for the two monotonic branches of the cosine function over a full period. This result is precisely the U-shaped Beta distribution $B(0.5, 0.5)$. The analytical prediction for this U-shaped distribution is numerically validated by the data presented in Fig.~\ref{fig_currentdensity}(d), which we elaborate below.

% === Figure 3 ===

\subsection{Numerical validation}
Figures~\ref{fig_mainresult}(c) and \ref{fig_mainresult}(d) present a comparison between our theoretical predictions and numerical calculations on the average conductance and conductance fluctuation. The analytical fits, based on Eq.~(\ref{eq_Gdis_analy}) and Eq.~(\ref{eq_sigmaG_analy}), are represented by dashed curves. The excellent agreement between the analytical and numerical results at weak and moderate disorder strengths justifies the theoretical framework developed in the preceding subsection.

The statistical distribution of $G$ serves as a unique fingerprint for quantum transport. To determine this numerically, the conductance range is partitioned into a discrete mesh $G_i$ with a bin-width $h=0.01$. The probability distribution $P_G(G_i)$ is then calculated as:
\begin{align}
P_G(G_i) = \frac{\mathcal{N}(G_i \leq G < G_i + h)}{h \mathcal{N}_{total}}
\end{align}
where $\mathcal{N}(G_i \leq G < G_i + h)$ denotes the number of conductance realizations falling within the interval $[G_i, G_i + h)$ across a total of $\mathcal{N}_{total}$ disorder configurations. To numerically verify Eq.~(\ref{eq_PG_Gauss}) and Eq.~(\ref{eq_PG_U}), we compare the conductance distributions at weak ($W = 0.1 E_0$) and moderate ($W =2.2 E_0$) disorder in Fig.~\ref{fig_currentdensity}(a) and (d), respectively. At weak disorder, the distribution follows a Gaussian profile, while at moderate disorder which belongs to the PAR, it adopts a U-shaped beta-function. Both cases show excellent agreement with our theoretical derivations (see the red dashed curves).
 
To gain microscopic insight into mode scattering at the disordered DW, we plot the spatial distribution of the current density ${\bf J}_{\bf i}$ in Fig.~\ref{fig_currentdensity}(b, c, e, f). These distributions reveal the electron partitioning process: electrons injected from the left electrode ($z<0$) via the THSs split into TESs at the DW interface. After crossing the DW, they are either transmitted to the right electrode or backscattered to the left. At weak and moderate disorder, the 3D current distribution remains hollow, indicating that the THSs and TESs retain their 1D propagating nature with nearly unaffected amplitudes. This confirms that the TESs at the DW edge maintain their unidirectional topological properties, justifying the assumptions used in our theoretical derivation.

\section{Theory of Mode-mixing}
\label{sec_mode_mixing}

\subsection{Derivations of conductance and conductance-fluctuation in the MMR}

In this subsection we develop the theory for the MMR occurring at strong disorder, where the TESs get completely mixed. In this regime, the $\sigma_G$-plateau collapses, the phase-averaging mechanism breaks down, and the analytical formula in Eq.~(\ref{eq_sigmaG_analy}) no longer applies [see Fig.~\ref{fig_mainresult}(d)]. Instead, we resort to the mode-mixing mechanism, which can be understood from two perspectives. 

First, according to RMT for a quantum chaotic cavity with broken time-reversal symmetry~\cite{AbaninQuantized2007, BarangeriMesoscopic1994, SavinShot2006, BeenakkerRandom1997}, the conductance average $\langle G \rangle$ is 
\begin{align}
\label{eq_G_RMT}
\langle G \rangle =  \frac{ N_1 N_2}{N_1 + N_2},
\end{align}
and the conductance variance is 
\begin{align}
\label{eq_varG}
{\rm Var}(G) = \frac{ (N_1 N_2)^2 } {(N_1 + N_2)^2 [(N_1 +  N_2 )^2 -1 ] },
\end{align}
where $N_{1(2)}$ is the number of channels of the left(right) electrode. In our system, $N_1 = N_2 = 1$, yielding $\langle G \rangle =1/2$, ${\rm Var}(G) = 1/12$ and $\sigma_G = {1}/{\sqrt{12}} \approx 0.289$. This result perfectly matches the second $\sigma_G$-plateau observed in Fig.~\ref{fig_mainresult}(d). 

Another explanation lies in the hypothesis of a uniform conductance distribution in the complete mode mixing case, as proposed in Ref.~\cite{LongScaling2025}:
\begin{equation}
 P_G(G) = 1, \quad G \in [0, 1].
\end{equation}
This uniform distribution yields an expectation value of $\langle G \rangle = 0.5$ and predicts a universal limit for conductance fluctuations:
\begin{equation}
 \sigma_G = \sqrt{\int_0^1 (G - 0.5)^2 dG} = \frac{1}{\sqrt{12}}.
\end{equation}
This result is identical to the RMT one. The excellent agreement between this derivation and the saturation plateau observed in Fig.~\ref{fig_mainresult}(d) confirms that the transition of $\sigma_G$ from 0.35 to 0.29 represents a definitive crossover from the PAR to MMR.

\subsection{Numerical validation}
Figure~\ref{fig_currentdensity}(g) displays the numerically calculated probability distribution of the conductance $G$ at strong disorder ($W=9 E_0$). In this regime, the transport resides in the MMR. The distribution is nearly uniform within $[0, 1]$, in excellent agreement with the analytical prediction for $P_G$ (see red dashed line).

In Fig.~\ref{fig_currentdensity}(h, i), we plot the spatial distribution of the current density ${\bf J}_{\bf i}$ at the DW. Under strong disorder ($W=9E_0$), the 1D nature of the transport is destroyed as pronounced scattering occurs between the THSs, the TESs, and the 3D bulk states. The current is no longer confined to the hinges or the DW interface; instead, it spreads predominantly into the bulk, forming a ``diffusive cloud". This provides direct evidence that strong disorder destroys topological protection and induces strong mixing between the upper-right and lower-left arms of the TESs. The unidirectional propagating properties of the THSs/TESs are thus broken and the PAR breaks down. 

% === Figure 4 ===
\begin{figure}[t]
  \centering
  \includegraphics[width=0.48\textwidth]{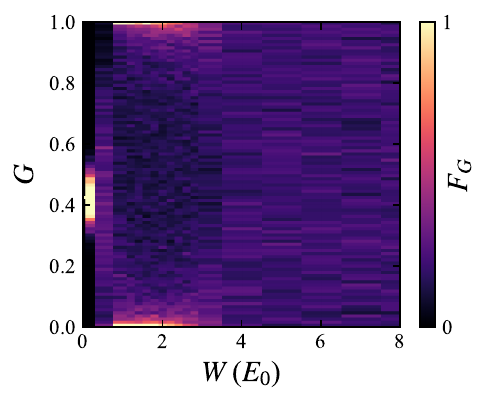} 
  \caption{\label{fig_global_evolution}
  Global evolution of transport regimes driven by disorder. The colormap illustrates the integrated probability $F_G$ of the conductance as a function of the disorder strength $W$. The magnetic flux is fixed at $\Phi = 4\pi/9$, the Fermi energy is $E_F = 0.02 E_0$, and the statistics for each $W$ are obtained from an ensemble of 3000 disorder configurations. The bin width is $h=0.01$.
  }
\end{figure}

In Fig.~\ref{fig_global_evolution} we present the global evolution of the conductance distribution. To limit the exhibition range into $[0, 1]$, we define the probability $F_G \equiv \int_{G}^{G+h} P_G(G_{\rm dis}) dG_{\rm dis}$ for a given bin-width $h$. The evolution follows four distinct stages: (1) Clean limit ($W \sim 0$): the probability $F_G$ exhibits a Dirac-delta profile. (2) Weak disorder: $F_G$ transitions into a Gaussian distribution, with its width broadening as $W$ increases. (3) PAR: As $W$ further increases, $F_G$ evolves into a U-shaped beta-function, signaling the randomization of the phase difference. (4) MMR: Under strong disorder, the distribution ultimately becomes uniform within $[0, 1]$, consistent with the complete mixing of transport channels.

\section{Fano factor} 
\label{sec_Fano}

The average-conductance as well as the conductance fluctuation can both be experimentally observed by varying the disorder configuration, or alternatively, by varying the chemical potential or magnetic flux~\cite{Lee1985Universal, Wang2016Universal} while fixing the disorder configuration. Here, we propose another way of measuring the second-order cumulants of conductance---the Fano factor---which is experimentally more feasible~\cite{KobayashiShot2021, MatsuoEdge2015, KumadaShot2015} for self-averaging but still phase-coherent transport systems. For instance, in time-dependent fluctuating backgate or electrostatic potentials, the time-average is equivalent to the ensemble average. While the conductance fluctuation is washed out by self-averaging, the Fano factor can still be extracted from the shot-noise measurement. We show that the Fano factor is also an important signal for identifying the PAR or MMR.

The Fano factor measures the ratio between the actual shot-noise power and the Poisson-noise power. In the framework of the Landauer-Büttiker formalism, the Fano factor at zero temperature is defined as~\cite{KobayashiShot2021}: 
$ F = {\sum_{n} \langle T_n(1-T_n)\rangle} / {\sum_n \langle T_n \rangle}$,
where $T_n$ is the transmission eigenvalue for the $n$-th individual channel. In our case, since there is only one incoming/reflecting channel and $G=T$ (omitting the units), the Fano factor is expressed in terms of the conductance $G$:
\begin{align}
F = {\langle G (1-G) \rangle }/ {\langle G \rangle }		.
\end{align}
 For the PAR, the Fano factor is calculated as:
\begin{align}
\label{eq_fano_phase}
F = \frac{1} {\langle G \rangle} \int_0^{2 \pi} P_{\Phi} G(\Phi_{\rm eff}) [1-G(\Phi_{\rm eff})] {\rm d} \Phi_{\rm eff} = \frac{1}{4}.
\end{align}
For the MMR, the Fano factor is
\begin{align}
\label{eq_fano_mode}
F = \frac{1} {\langle G \rangle} \int_0^1 P_G G (1-G) {\rm d} G = \frac{1}{3}.
\end{align}
In Eq.~(\ref{eq_fano_phase}) and Eq.~(\ref{eq_fano_mode}), we have utilized the uniform distributions of the effective flux $\Phi_{\rm eff}$ and the conductance $G$, respectively. The Fano factor in the MMR can also be obtained from RMT~\cite{AbaninQuantized2007, BarangeriMesoscopic1994, SavinShot2006, BeenakkerRandom1997}:
\begin{align}
F= \frac{N_1 N_2} {(N_1 + N_2)^2 -1},
\end{align}
where in our case of $N_1=N_2 =1$ we get $F = 1/3$. This value is identical to the Fano factor $F$ found in metallic diffusive conductors~\cite{BeenakkerRandom1997, KobayashiShot2021}.

% === Figure 5 ===
\begin{figure}[t]
    \centering
    \includegraphics[width=0.95\columnwidth]{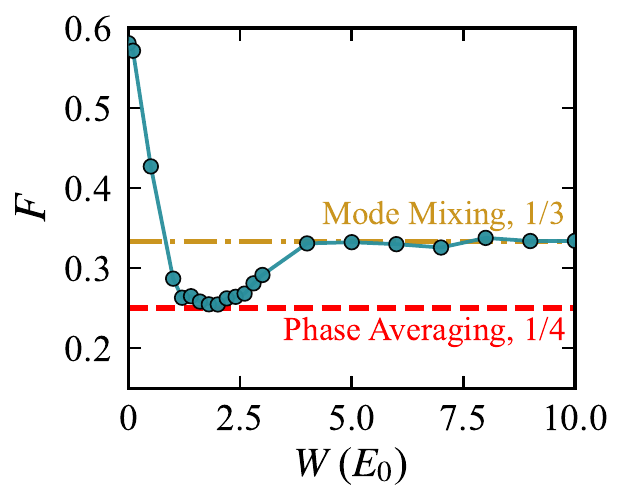} 
    \caption{\label{fig_fano}
Fano factor $F$ as a function of the disorder strength $W$. The red and brown dashed curve at $F=1/4$ and $F=1/3$ denote the theoretical Fano factor values in the PAR and MMR, respectively. Here the magnetic flux is fixed at $\Phi = 4\pi/9$, the Fermi energy is $E_F = 0.02 E_0$, and the results are averaged over 3000 disorder configurations.
   }
\end{figure}

In Fig.~\ref{fig_fano} we plot the Fano factor $F$ as a function of the disorder strength $W$. At $W=0$, despite the absence of disorder, mode partitioning still occurs due to the AB effect at the DW, resulting in a nonzero $F$. As $W$ increases, disorder scattering becomes dominant and the transport evolves into the PAR. Consequently, the Fano factor $F$ decreases and reaches the first plateau of 1/4 at $W \approx 2 E_0$, which is the hallmark of the PAR. This plateau persists over a short range of $W$; upon further increasing $W$, $F$ increases to a second plateau at 1/3, signaling the onset of MMR. These numerical results are in perfect agreement with the theoretical predictions in Eq.~(\ref{eq_fano_phase}) and Eq.~(\ref{eq_fano_mode}), demonstrating that the Fano factor serves as a robust signal for observing the crossover from the PAR to MMR at disordered DWs. Notably, similar evolution behavior of the Fano factor is observed in graphene quantum point contacts~\cite{Vilvanathan2026Landau}, though with a different physical mechanism.

% === Figure 6 ===
\begin{figure}[t!]
  \centering
  \includegraphics[width=\columnwidth]{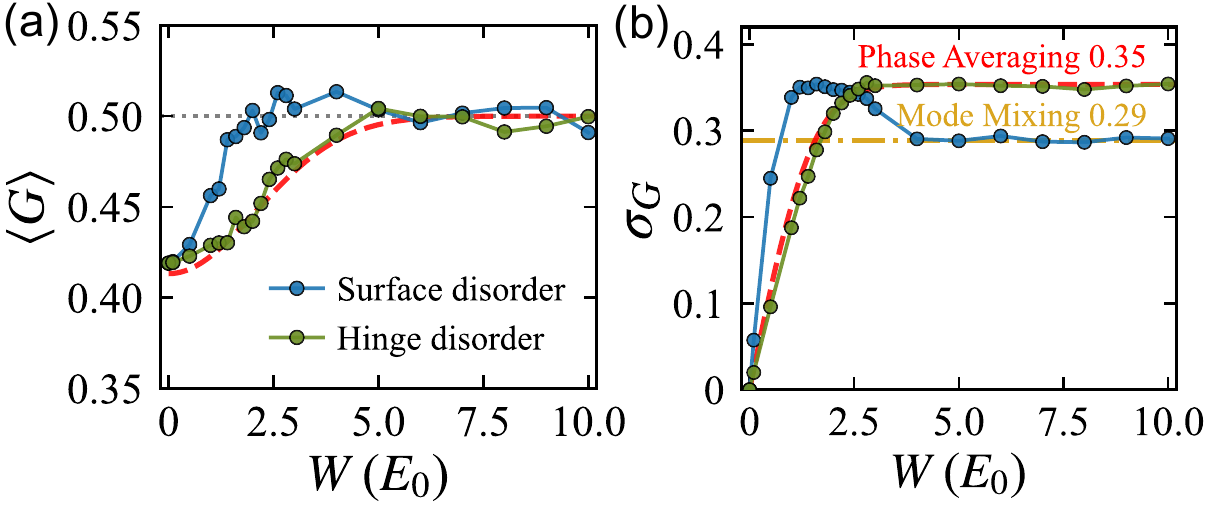} 
  \caption{\label{fig_surface_hinge}
Effects of surface disorder and hinge disorder on the transport results. (a) Ensemble-averaged conductance $\langle G \rangle$ and (b) conductance fluctuation $\sigma_G$ as functions of disorder strength $W$ for surface disorder and hinge disorder. The red dashed curve denote the theoretical fitting for the hinge disorder, yielding a fitting parameter $\chi_{\text{hinge}} \approx 0.117$. The magnetic flux is fixed at $\Phi = 4\pi/9$, the Fermi energy is $E_F = 0.02 E_0$, and the results for each $W$ are averaged over 3000 disorder configurations. 
  }
\end{figure}

\section{Dependence on the spatial location of disorder } 
\label{sec_disorder_location}
In the previous sections, we assumed that disorder exists throughout the entire bulk of the central region (bulk disorder case). To examine whether other types of disorder---such as surface disorder, where disorder exists exclusively on the surface of the central region, or hinge disorder, where disorder exists only on the four hinges of the central region---can induce a crossover between the two transport regimes, in this section we vary the spatial location of the disorder and study their influence on transport.

Figure~\ref{fig_surface_hinge}(a) shows the average conductance $\langle G \rangle$ versus $W$ for the surface disorder (blue curve) and hinge disorder (red curve), with the corresponding conductance fluctuations $\sigma_G$ plotted in Fig.~\ref{fig_surface_hinge}(b). As shown, $\langle G \rangle$ for both cases increases with $W$ and eventually saturates at the half-quantized CP of 0.5. However, the conductance fluctuation exhibits distinct behaviors. For surface disorder, it displays a double-plateau structure at approximately 0.35 and 0.29, consistent with the bulk disorder results in Fig.~\ref{fig_mainresult}(d). This indicates both the PAR and MMR exist for surface disorder, and the mode-mixing at strong disorder should be mediated by the surface states [see yellow curves in Fig.~\ref{fig_mainresult}(a)]. In contrast, $\sigma_G$ under hinge disorder shows only a single 0.35 plateau and does not collapse, implying that only the PAR occurs and remains stable against disorder strengths. Furthermore, $\langle G \rangle$ and $\sigma_G$ for hinge disorder approach their respective plateaus more slowly than those for surface disorder due to the reduced disordered area. This is validated by the curve fitting using Eq~(\ref{eq_Gdis_analy}) and Eq.~(\ref{eq_sigmaG_analy}), which yields a modification factor $\chi \approx 0.117$, much smaller than the value $\chi \approx 0.733$ obtained for bulk disorder. The Fano factors for surface and hinge disorder are also calculated, which show a double-step plateau (1/4 and 1/3) and a single plateau (1/4), respectively (results not shown). In Appendix~\ref{sec_surface_hinge_statistics} we provide the conductance statistics and local current density distributions under strong disorder. The results are in perfect agreement with those found for the PAR and MMR of bulk disorder. So to conclude, observing the crossover from PAR to MMR requires disorder to exist either in the bulk or on the surfaces, while hinge disorder only supports the existence of PAR.

\section{Conclusion}
\label{sec_conclusion}
In conclusion, by investigating quantum transport across a disordered magnetic DW in a 3D SOTI nanowire, we have uncovered a disorder-induced crossover from PAR to MMR. At moderate disorder, the TESs surrounding the DW retain their unidirectional-propagation property, allowing for the randomization of the dynamical phase difference. This leads to a PAR characterized by a half-quantized average conductance at $\langle G \rangle = 0.5$ and a specific conductance fluctuation plateau at $\sigma_G \approx 0.35$. As the disorder strength increases, the unidirectional nature of the TESs is destroyed, inducing strong inter-mode scattering between the interference paths. This leads to the emergence of the MMR, which maintains the half-quantized average conductance while exhibiting a distinct fluctuation plateau at $\sigma_G \approx 0.29$.

To elucidate these results, we developed analytical theories for both regimes that demonstrate excellent agreement with the behaviors of $\langle G \rangle$ and $\sigma_G$. Our findings are further corroborated by large-scale numerical simulations on the probability distributions of conductance and spatial current density profiles. Furthermore, the Fano factor associated with shot-noise measurements is calculated, which exhibits a similar two-step evolution transitioning from a 1/4 plateau in the PAR to a 1/3 plateau in the MMR. This offers a clear metric for experimental verifications on the PAR-MMR crossover. Finally, the influence of the spatial location of disorder on transport is investigated, which reveals that both bulk and surface disorder facilitate this crossover, while hinge-only disorder exclusively supports the PAR.

Our work thus proposes the magnetic DW in a 3D SOTI as a unified platform for studying the interplay between phase-averaging and mode-mixing physics. By identifying conductance fluctuations of 0.35 and 0.29, or the Fano factor of 1/4 and 1/3 as distinct statistical fingerprints, we provide precise quantitative criteria for experimentally distinguishing between the two regimes. The results also suggest disorder-engineering as a powerful route for controlling electronic transport across magnetic DWs, offering potential applications in DW-based topological and spintronic devices.

\section*{Acknowledgments}
We thank Jiayin Gu for valuable discussions. This work is supported by the National Natural Science Foundation of China under Grants No. 12304070.

\section*{Data Availability} 
The simulation code that supports the findings of this article is available upon request.

\appendix

\section{Derivation on the explicit expressions of $\langle G \rangle$ and $\sigma_G$}
\label{app_Gaussian_integral}
To obtain Eq.~(\ref{eq_Gdis_analy}), we calculate the integral:
\begin{align}
\langle G_{\rm dis}(\Phi)\rangle &= \int_{-\infty}^{\infty} G (\Phi + \Delta \varphi) P_{\varphi}(\Delta \varphi)\, d\Delta \varphi \notag \\
&= \int_{-\infty}^{\infty} \left[\frac{1}{2}-\frac{1}{2}\cos(\Phi+\Delta \varphi)\right] \notag \\
&\quad \times \frac{1}{\sqrt{2\pi \sigma_{\varphi}^{2}}} e^{-\frac{ (\Delta \varphi)^{2}}{2\sigma_{\varphi}^{2}}} \, d\Delta \varphi\,.
\end{align}
By substituting $t = \Delta \varphi + \Phi$, we get:
\begin{align}
\langle G_{\rm dis}(\Phi)\rangle &= \frac{1}{2} - \frac{1}{ 2 \sqrt{2\pi \sigma_{\varphi}^{2}}} \int_{-\infty}^{\infty} \cos(t) e^{-\frac{ (t-\Phi)^{2}}{2\sigma_{\varphi}^{2}}} dt \notag \\
&= \frac{1}{2} - \frac{1}{ 2 \sqrt{2\pi \sigma_{\varphi}^{2}}} {\rm Re} \left\{ \int_{-\infty}^{\infty} e^{i t} e^{-\frac{ (t-\Phi)^{2}}{2\sigma_{\varphi}^{2}}} \, dt \right\} \notag \\
&= \frac{1}{2} - \frac{1}{2}e^{-\frac{ \sigma_{\varphi}^{2}}{2}}\cos\Phi \notag \\
&= \frac{1}{2} - \frac{1}{2}e^{-\frac{N}{6}\widetilde{W}^2}\cos\Phi. \label{eq_b1_final}
\end{align}
Here in the last line we used $\sigma_{\varphi} = \frac{\sqrt{3N}}{3} \widetilde{W}$.

To obtain Eq.~(\ref{eq_sigmaG_analy}), we calculate:
\begin{align}
{\rm Var}(G) &= \int_{-\infty}^{\infty} \big[G(\Phi + \Delta \varphi )-\langle G_{\rm dis}(\Phi) \rangle \big]^{2} P_{\varphi}( \Delta\varphi ) \, d\Delta\varphi \notag \\
&= \frac{1}{4} \int_{-\infty}^{\infty} \left[ \cos(\Phi+ \Delta \varphi) - e^{-\frac{N}{6}\widetilde{W}^{2}}\cos\Phi \right]^{2} \notag \\
&\quad \times \frac{1}{\sqrt{2\pi \sigma_{\varphi}^{2}}} e^{-\frac{(\Delta \varphi)^{2}}{2 \sigma_{\varphi}^{2}}} d\Delta \varphi \notag \\
&= \frac{1}{4 \sqrt{2\pi \sigma_{\varphi}^{2}}} \int_{-\infty}^{+\infty} \left[ \cos^{2}(t)-2A\cos(t)+A^{2} \right] \notag \\
&\quad \times e^{-\frac{(t-\Phi)^{2}}{2\sigma_{\varphi}^{2}}} dt
\end{align}
where we have denoted $t=\Delta \varphi + \Phi$ and $A = e^{-\frac{N}{6} \widetilde{W}^2} \cos\Phi$. Replacing $\cos^2(t)$ with $[\cos(2t)+1]/2$ gives
\begin{align}
{\rm Var}(G) &= \frac{1}{8 \sqrt{2\pi \sigma_{\varphi}^{2}}} {\rm Re} \left\{ \int_{-\infty}^{+\infty}e^{2 i t } e^{-\frac{(t- \Phi)^{2}}{2 \sigma_{\varphi}^{2}}} dt  \right\}   +\frac{1}{8}    -\frac{A^{2}}{4} 	\notag \\
&= \frac{\cos(2\Phi)}{8 }  e^{-2\sigma_{\varphi}^{2}}  +\frac{1}{8}    -\frac{A^{2}}{4} 	\notag \\
&=  \frac{1}{8}[ 1 - \cos(2\Phi)e^{-\frac{N}{3}\widetilde{W}^{2}} ]\cdot( 1 - e^{-\frac{N}{3}\widetilde{W}^{2}} )	. 
\end{align} \\

% === Figure 7 ===
\begin{figure*}[t]
  \centering
  \includegraphics[width=0.95\textwidth]{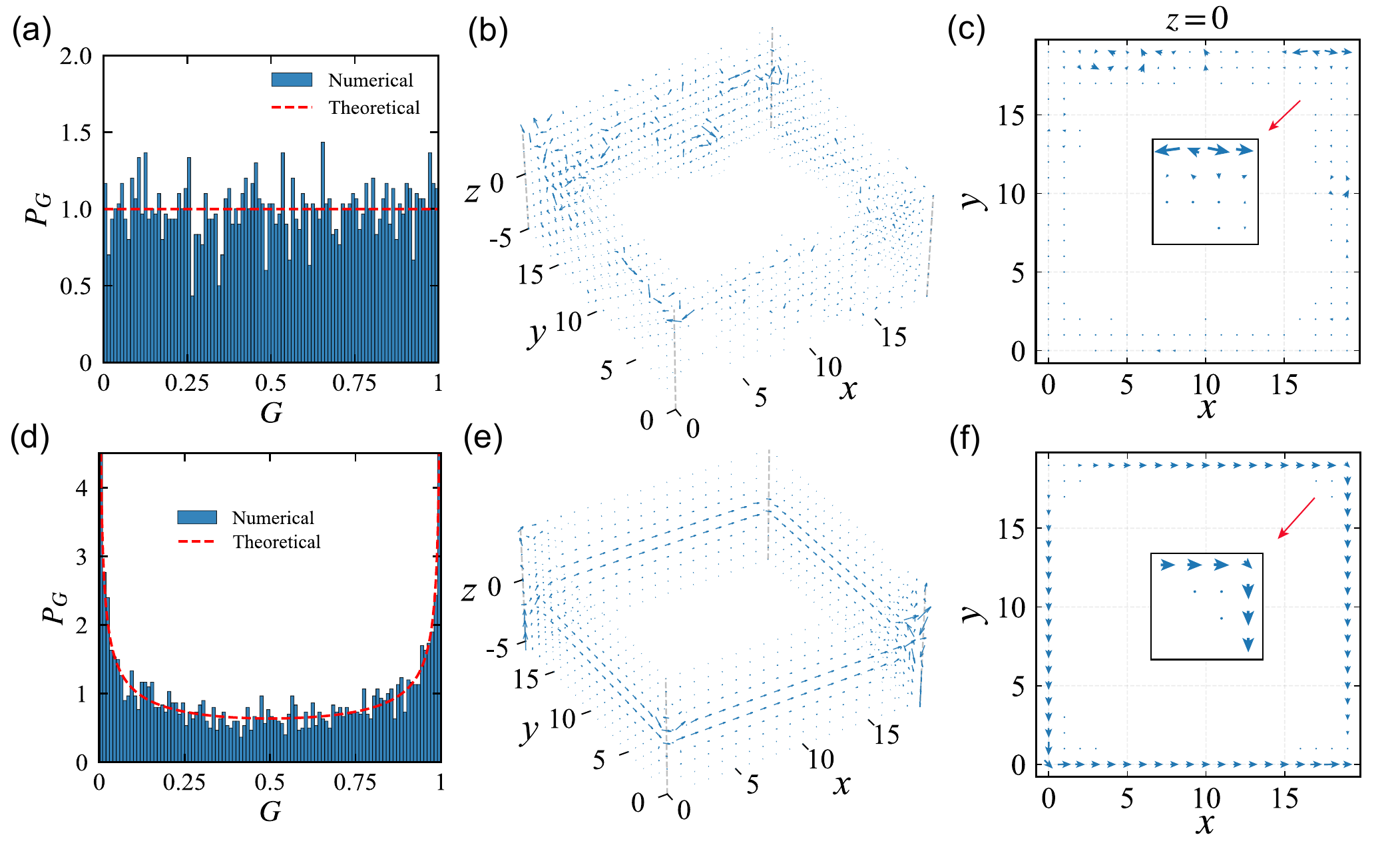} 
  \caption{\label{fig_current_surface_hinge}
  Comparison of microscopic transport features under changing spatial locations of disorder at strong disorder ($W=9.0 E_0$). (a)--(c) Surface-only disorder scenario: (a) Probability distribution $P_G$; (b) 3D visualization of the ``hollow'' current density distribution; (c) Cross-sectional view of current at $z=0$. (d)--(f) Hinge-only disorder scenario: (d) Probability distribution $P_G$; (e) 3D visualization of the current density; (f) Cross-sectional view at $z=0$. The red dashed curves in (a) and (d) show the analytical results of $P_G$. The magnetic flux is fixed at $\Phi = 4\pi/9$, the Fermi energy is $E_F = 0.02 E_0$, and the statistical distributions of $P_G$ in (a) and (d) are each obtained from an ensemble of 3000 disorder configurations with a bin-width of $h=0.01$. The insets in (c) and (f) provide magnified views of the local current density. 
  }
\end{figure*}

\section{Conductance statistics and local current density distribution for surface-disorder and hinge-disorder } 
\label{sec_surface_hinge_statistics}
We present the numerically calculated probability distribution of $G$ for surface disorder and hinge disorder in Fig.~\ref{fig_current_surface_hinge}(a) and (d), respectively, with the disorder-strength set to be strong ($W=9 E_0$). As shown, the distribution is uniform for surface disorder and U-shaped for hinge disorder, conforming with the theories of the MMR and PAR, respectively.

To gain further insight into the mechanisms underlying these distinct transport behaviors, we plot the current density distribution at the DW for both surface and hinge disorder in Fig.~\ref{fig_current_surface_hinge}(b,c,e,f). In the case of surface disorder, scattering occurs between the TESs and the 2D surface states, which mediates the mode-mixing between the upper-right and lower-left arms of the TESs. This is evident in Fig.~\ref{fig_current_surface_hinge}(b), where the current exhibits slight penetration into the bulk region and undergoes direction changes induced by disorder, indicating a breakdown of the unidirectional propagation characteristic of the TESs. By contrast, for hinge-only disorder, scattering is confined to the hinges. While the current direction at the hinges is altered [see Fig.~\ref{fig_current_surface_hinge}(e)], the propagation of the TESs remains largely unaffected due to the absence of disorder along their paths. Since the disorder is restricted to a central region of finite length $L$, the current on the THSs, despite being scattered, must ultimately flow into the TESs. The enclosed interference paths formed by the TESs remain intact; consequently, the primary effect of the disorder is the introduction of random dynamical phases for the two interference arms, regardless of the disorder strength. In this case, the phase-averaging theory remains applicable across all disorder strengths, resulting in a robust PAR. \\

\bibliography{myrefs}         

\end{document}